# Blockchain-Driven Research in Personality-Based Distributed Pair Programming

Marcel Valovy*
*Department of Information Technologies*
*Prague University of Economics and Business*
Prague, Czech Republic
https://orcid.org/0000-0001-7074-0918
*Corresponding author

Alena Buchalcevova
*Department of Information Technologies*
*Prague University of Economics and Business*
Prague, Czech Republic
https://orcid.org/0000-0002-8185-5208

*Abstract*—This study aims to integrate blockchain technology into personality-based pair programming research to enhance its generalizability and adaptability by offering built-in continuous, reproducible, and transparent research. In the developing Role-Optimization Motivation Alignment (ROMA) framework, human/AI programming roles align with individual Big Five personality traits, optimizing individual motivation and team productivity in Very Small Entities and undergraduate courses. Twelve quasi-experimental sessions were conducted to verify the personality-based pair programming in distributed settings. A mixed-methods approach was employed, combining intrinsic motivation inventories and qualitative insights. Data were stored transparently on the Solana blockchain, and a web-based application was developed in Rust and TypeScript languages to facilitate partner matching based on ROMA suggestions, expertise, and availability. The results suggest that blockchain can enhance research generalizability, reproducibility, and transparency, while ROMA can increase individual motivation and team performance. Future work can focus on integrating smart contracts for transparent and versioned data analysis.

## I. INTRODUCTION

This project seeks to establish a Blockchain-Driven Continuous Generalization and Adaptation (BDCGA) framework to facilitate transparent and reproducible research, enhancing socio-cultural validation and adaptation, and to evaluate its efficacy by integrating it into current personality-based pair programming research. In a world where emerging technologies perpetually create new programming positions, roles, and environments, BDCGA maintains findings current and adheres to research ethics, such as reproducibility, by consistently integrating new empirical data through a tamper-proof ledger on the Solana blockchain.

### A. Related Work

Based on behavioral software engineering research, the recently developed *ROMA* framework [1] focuses on how individual personality factors can affect programming motivation and productivity [2-4]. Assigning roles like Pilot and Navigator in traditional pair programming can improve code quality, teamwork, and knowledge exchange [5,6]. However, how effectively a person succeeds in these jobs is greatly influenced by individual differences, especially personality traits like neuroticism, extraversion, and openness [7-9]. The ROMA framework incorporates the Big Five personality traits to enhance the role assignment process. According to Self-determination Theory, it assigns individuals to programming roles where they thrive, meeting their basic psychological needs and maximizing their well-being [10].

*Previous blockchain and distributed systems research* has explored blockchain technology's potential for decentralizing and securing personal data [11]. More recently, blockchain applications have expanded into areas including research reproducibility and supply chain management [12,13]. Building on this research, this study uses Solana's blockchain to process and store empirical data collected through opt-in in the deployed application, providing research replicability and transparency. It encourages distributed and versioned data analysis in future studies through custom Solana Program Libraries (SPLs) [14].

### B. Context

This study extends a series of mixed-method experiments in personality-based human/AI pair programming that were carried out in professional and student settings between 2021 and 2024 [3,4,15]. The ROMA framework, which presently targets VSEs and educational environments, was developed based on these research findings [1]. Although this research on role optimization is especially pertinent to situations with limited resources, the ROMA framework can be adapted to a variety of software development contexts. Academic teams and VSEs frequently have small teams and need team members to wear multiple hats. In such contexts, role optimization and motivation alignment are crucial in order to maximize collaborative performance and team productivity [16].

### C. Our Contribution

The presented study on BDCGA and ROMA is guided by:

**RQ1:** Can personality-based role optimization be used to align intrinsic motivation in pair programming?

© 2024. This preprint is the authors' version of the work. The final version will be published in the Proceedings of the Asia Conference on Information Engineering (ACIE 2025), IEEE Xplore. This work was supported by an internal grant funding scheme (F4/61/2023) administered by the Prague University of Economics and Business.

**RQ2:** How can blockchain technology improve empirical software engineering reproducibility & transparency?

**RQ3:** How can personality-based role optimizations be adapted to distributed and very small entity environments?

These research questions are pivotal in guiding the research into the theoretical and practical use of blockchain technology in empirical software engineering and laying the groundwork for the generalizability and adaptability of the ROMA framework. This research also explores artificial intelligence (AI) and smart contracts (SPL) integration as future prospects.

## II. ROLE-OPTIMIZATION MOTIVATION ALIGNMENT

We illustrate the ROMA framework's potential for personality-based role optimization by associating personality clusters characterized by their predominant Big Five personality trait with their preferred roles, where they reach the highest average intrinsic motivation, described in detail in *Table I*.

TABLE I. PERSONALITY CLUSTERS AND MOTIVATIONAL FOCUS IN THE ROLE-OPTIMIZATION MOTIVATION ALIGNMENT (ROMA) FRAMEWORK

| Cluster | Role Optimization | | | |
|---|---|---|---|---|
| | *Personality Traits* | *Role* | *Key Role Strengths* | *Motivation Focus* |
| Cluster 1 | Openness | Pilot | - Creativity and curiosity in problem solving<br>- Openness to partner's ideas and conceptual thinking | Engages well in novel tasks and **exploratory** problem-solving. Thrives in environments encouraging **expression** and idea generation. |
| Cluster 2 | Extraversion & Agreeableness | Navigator | - Effective communication and guidance<br>- Collaborative nature ensures team cohesion | Excels in roles requiring social **interaction** and cooperation. Motivates by interactively **guiding** coding, quality, structure. |
| Cluster 3 | Neuroticism & Introversion | Solo | - Focused on structured, predictable tasks<br>- Independent and detail-oriented | Thrives in independent work, **without the pressure** of continuous social interaction. Motivated by **structured**, well-defined tasks with less social interaction. |

By clustering participants based on their Big Five traits, the ROMA framework effectively assigns roles that align with an individual's strengths—e.g., those high in Openness are strongly associated with divergent thinking and the ability to generate multiple unusual and creative solutions to problems [8].

Based on this, *two hypotheses* were proposed:

- **H1** and **H1-Cor**: *Motivation across pair programming roles differs, and–corollary–the difference in individual average intrinsic motivation across roles is consistent.*

- **H2:** *Individuals of high Openness prefer the Pilot role.*

*Figure 1* visualizes how the ROMA framework integrates personality-based pair programming into software engineering methodologies, particularly enhancing the Project Management (PM) and Software Implementation (SI) processes of agile Basic Profile in ISO/IEC 29110 standard series for VSEs [16,17].

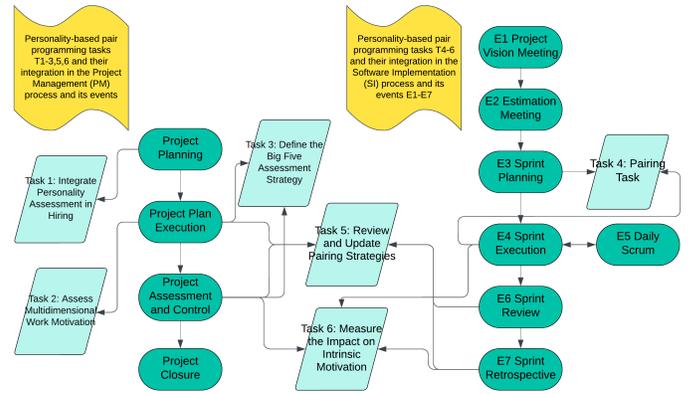

Fig. 1. Personality-Based Pair Programming tasks for the PM and SI processes of the ISO/IEC 29110 Software Basic Profile and Agile Guidelines.

The ROMA extension for VSEs comprises six tasks. This study proposes an application to address *Tasks 1, 3, 4, and 6* by integrating personality assessments into project initiation stages, providing personality-based strategies for role assignment, facilitating team pairings, and measuring the role alignment impact on intrinsic motivation, ensuring optimal productivity. In academic and professional settings, ROMA aims to foster better teamwork, enhance student engagement, and aid independent professionals in finding well-matched collaborators.

## III. METHODOLOGY

### A. System Architecture and Technological Stack

To support the ROMA framework, we developed a robust system architecture integrating the *Solana blockchain*, *Rust backend*, and a *React/TypeScript frontend*. It facilitates pairing sessions and seamlessly collects and stores experimental data.

*1) Backend with Rust and Solana Integration:* The backend was implemented in Rust, leveraging its high-performance and safety capabilities. It was integrated with the Solana blockchain, which also employs Rust, ensuring real-time experimental data handling due to Solana's high throughput and low transaction costs. The system is currently deployed on *Solana Devnet.*

*2) Data Storage and Security:* Solana transaction memos were used to store experimental data on the Solana blockchain, as illustrated in *Fig. 2*. Publicly available blockchain explorers and dedicated dashboards can be used to transparently visualize session outcomes. To ensure *GDPR* compliance, user data were recorded using *SHA-256 hashed IDs,* preserving anonymity.

*3) Frontend Application with React and TypeScript:* The frontend was developed using *React v18* and TypeScript, offering an intuitive interface for user registration, personality assessment, and session scheduling. The frontend communicates securely with the backend via *JWT tokens* for authentication and *secure API calls* for real-time session coordination. Participants can view their scheduled pair programming sessions and recommended role assignments through integrated calendars. The frontend also supports personalized feedback collection through the Intrinsic Motivation Inventory form submission.

*4) Screen and Communication Tools:* To facilitate distributed pair programming, the system utilized tools like

IntelliJ's Code With Me and Visual Studio's Live Share, which enabled screen sharing, remote control, and audiovisual communication. Participants shared links to sessions through the frontend's calendars to collaborate in real-time, regardless of geographic location, closely mirroring in-person experience.

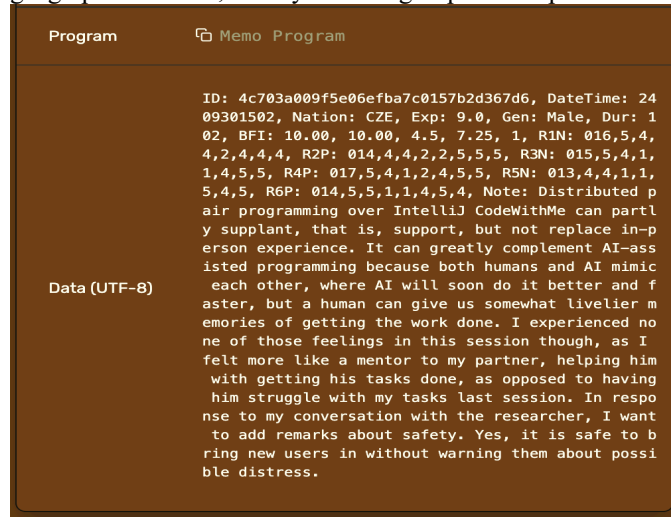

Fig. 2. Solana Blockchain transaction memo storing a single session data containing motivation inventories for six experimental rounds of personality-based distributed pair programming and open-ended feedback.

### B. Quasi-Experimental Design

*1) Participants.* The pilot study involved four participants drawn from the VSE talent pool. Informed consent was provided, and the study followed strict ethical guidelines.

*2) Procedure.* Participants simulated a quasi-experiment by engaging in *two pair programming sessions* and *one solo session*. Each session was structured with six rounds of varying lengths (12-18 minutes, $\bar{x} = 14.97$, $s = 1.28$). The participants worked on tasks of their choice, switching between roles. The preferred roles with suggested double time were assigned by the ROMA framework, using data from the Big Five Inventory (BFI-10) personality assessment [18].

*3) Measures.* Quantitative and qualitative measures were employed to assess the effects of personality-based sessions:
- Quantitative: Motivation and engagement across roles were gauged after each round, where participants completed the Intrinsic Motivation Inventory's *Enjoyment/Interest subscale* [19].
- Qualitative: Participants shared insights on collaboration dynamics via open-ended feedback after each session.

### C. Quantitative and Qualitative Analysis

Quantitative data from twelve sessions were analyzed using *ANOVA* and *Kruskal-Wallis tests* to evaluate role-based differences in motivation. The consistency of motivational increases across roles was further tested using *Friedman rank sum* and *paired t-tests*. Future iterations of this system aim to streamline data processing and automate analysis using SPLs.

A *descriptive phenomenological approach* was conducted to articulate deeper meanings and subjective realities in distributed pair programming and solo programming contexts [20].

## IV. RESULTS

### A. Quantitative Findings

The study analyzed data from a convenient sample of four Czech VSE professionals who participated in two distributed pair programming sessions and one solo programming session each, divided into six rounds, yielding 72 analysis data points.

*1) Participants' Backgrounds:* The participants were three males and one female, ranging in experience from 1.5 to 9 years ($\bar{x} = 5.0$, $s = 3.43$). All participants exhibited high *Openness* as their predominant trait (*ROMA Cluster 1*). This homogeneity is likely due to the limited sample size and convenient sampling.

*2) Round Duration:* Across all sessions, the average round duration was *14.97 minutes* ($s = 1.28$, $min = 12$, $max = 18$), closely matching the recommended time of 15 minutes, because all participants preferred the same role. This consistency across participants and roles ensured that no specific role or participant caused significant deviations in session structure, maintaining experimental control.

*3) Personality Traits.* Linearly rescaled BFI-10 personality data to a *1-10 scale* revealed that participants exhibited high *Openness* ($\bar{x} = 9.25$, $s = 0.55$) and *Conscientiousness* ($\bar{x} = 8.88$, $s = 1.44$), impacting role preference. Other personality traits, including *Extraversion* ($\bar{x} = 4.38$, $s = 1.52$), *Agreeableness* ($\bar{x} = 5.22$, $s = 1.60$), and *Neuroticism* ($\bar{x} = 4.38$, $s = 2.35$), were moderate, impacting motivation in programming roles less.

### B. Statistical Significance Testing

Hypothesis **H1** was tested using *ANOVA*, which revealed a *near-significant* variance in motivation across roles ($F(2, 9) = 3.88$, $p = 0.061$), further *supported* by the *Kruskal-Wallis test* ($\chi^2(2) = 5.65$, $p = 0.059$). These results indicate that role differences in motivation approached statistical significance, limited by the small sample size.

The consistency of individual intrinsic motivation across roles (**H1-Cor**) was further validated using the *Friedman rank sum test* ($\chi^2(2) = 5.65$, $p = 6.14 \times 10^{-6}$). A *paired t-test* revealed a *significant* difference between maximum and minimum motivation scores, $t(3) = 4.21$, $p = 0.024$, with an average difference of 2.17 points (*95% CI:* 0.53, 3.81). These results *support* the conclusion that *personality influences motivational differences in specific roles*.

Motivation scores varied significantly across roles. Participants with high Openness were most motivated in the Pilot role ($\bar{x} = 8.45$, $s = 0.76$), followed by Navigator ($\bar{x} = 7.01$, $s = 0.63$), and least motivated in Solo ($\bar{x} = 6.87$, $s = 1.17$) (see Table II). These results support hypothesis **H2**.

TABLE II. INTRINSIC MOTIVATION OF ROMA CLUSTER 1 MEMBERS IN PILOT, NAVIGATOR, AND SOLO ROLES.

| Role | Intrinsic Motivation (scaled 1-10) | |
|---|---|---|
| | *Mean Motivation* | *Standard Deviation* |
| Pilot | 8.45 | 0.76 |
| Navigator | 7.01 | 0.63 |
| Solo | 6.87 | 1.17 |

*C. Qualitative Findings*

Participants' lived experiences during pair programming and solo sessions revealed novel insights into role engagement, learning, and human-human and human-AI collaborations.

Participants consistently reported greater *engagement and satisfaction* when acting as the *Pilot*. One participant (*xalp00*) remarked, *"I like working as a team, especially when I get to be the pilot because it's fun and I can do more. But when I'm the navigator, it's kind of boring since I just have to sit and watch."*

Comparing human pairing with *AI-assisted programming*, which naturally enriched their solo experiences, some participants appreciated the efficiency of AI but found *human interaction more fulfilling*. As participant *xtok06* noted, *"(...) both humans and AI mimic each other, where AI will soon do it better and faster, but a human can give us somewhat livelier memories of getting the work done,"* highlighting the irreplaceable social and creative aspects of human collaboration.

In *solo sessions*, participants enjoyed the focus but lacked the energy of pairing, as *xadk00* noted, *"The solo session gave me time to reflect, but it lacked the energy and dynamic flow that I enjoyed when pairing with a more experienced partner,"* marking the collaborative essence in motivation and learning.

There was consensus across participants that pair programming is *challenging*, especially with more experienced partners, *encouraging personal growth*. As *xadk00* remarked, *"I was pushed to think faster, ask better questions, and really sharpen my communication."* This feedback reflects the broader learning and motivational benefits of working collaboratively.

The considerable variance in *skill levels* required adjusting the *pace of work*. As *xika12* reflected, *"(...) she did mention that I was going a bit too fast when she was navigating. Interestingly, she was much more interactive when she was piloting,"* confirming that superior skills are desirable for navigators [4].

## V. Discussion

This study integrated the ROMA framework and blockchain technology for adaptable personality-based role optimizations. The results provided fresh insights for VSEs and empirical software engineering by matching programming roles with individual personality traits and evaluating the experimental effects on the transparent and programmable Solana blockchain.

*A. Role Optimization and Motivation Alignment*

The results addressed **RQ1** by quantitatively supporting hypotheses H1 and H2, indicating that programming roles can be optimized to align individual personality traits with intrinsic motivation. The *Pilot* role yielded the highest intrinsic motivation scores, particularly for participants exhibiting *Openness to Experience*, confirming previous research about creativity and exploration [3,4,8]. This suggests that the ROMA framework's role-personality alignment may positively impact motivation and performance, particularly if participants can form complementary pairing constellations and spend more time in their preferred roles. Future research should investigate this effect further, using larger and varied personality samples.

The study did, however, also reveal novel subtleties in *role engagement*. Despite being essential for fostering collaborative spirit, the Navigator function was seen as more passive, which led to lower motivation scores. This was particularly true for participants who had below-average extraversion and higher openness, which may have contributed to their desire for a more stimulating and dynamic role. These results somewhat deviate from earlier research, when the Navigator position was frequently regarded as inherently interactive [3,4]. The disparity might result from the distributed nature of pair programming sessions, where engagement may have decreased as a result of distant collaboration. In order to preserve engagement for all roles and avoid possible disengagement, future research should examine how remote environments might be optimized.

The *Solo* role lacked the dynamic flow and collaborative energy of pair programming, although providing a contemplative experience. The Solo position had the lowest motivation scores among participants, even with optional AI support. This suggests that the role's introspective nature may not fully meet basic psychological needs like collaborative situations do. Even though solo programming offered chances for reflection and concentration, it lacked the challenge and excitement of team-based activities, especially for people who do best in collaborative and interactive environments. This is consistent with earlier studies that found teamwork increased motivation and engagement more than working alone [6,7,10].

*B. Blockchain-Driven Research Reproducibility*

This study's use of the Solana blockchain to enable the durability, traceability, and transparency of experimental data is one of its distinctive contributions. We responded to **RQ2** and addressed the persistent problem of reproducibility in empirical research by recording session data on the blockchain [21]. Blockchain thus provides a workable answer to reproducibility in research by virtue of its permanent nature, which guarantees that the data can be independently validated by third parties [14].

*C. Distributed Personality-Based Pair Programming*

To answer **RQ3**, the proposed distributed pair programming application democratized access to diverse talent pools and created opportunities for independent partnerships. It aided the ROMA framework in satisfying basic psychological needs and raising optimal collaboration, challenges, and motivation in participants drawn from distributed VSE talent pools.

Using the ROMA distributed pair programming application, VSEs can optimize team performance and satisfaction. In educational settings, the ROMA application can help instructors assign roles that boost student engagement and collaboration. It can also benefit independent practitioners by matching them with compatible partners based on individual characteristics.

*D. Limitations*

Despite being exploratory in nature, this research has several limitations. The tiny sample size–there were only four participants, all of whom had high Openness scores–was a significant limitation. This homogeneity restricts the generalizability of the results across different personality profiles. To enhance the reliability of the findings, future studies should incorporate bigger and more varied participant samples.

The quasi-experimental design, while suitable for initial exploration, lacked the randomization rigor, which could help control for confounding variables. Additionally, the use of self-reported motivation inventories may have introduced response biases, potentially impacting the accuracy of the findings. To triangulate the findings, future research should utilize objective behavioral metrics, such as real-time performance tracking.

Furthermore, while Solana's blockchain technology added transparency, real-time accessibility, and durability to the data, its transaction memo limits and the complexity of storing rich datasets on-chain pose challenges. More robust blockchain solutions, such as smart contracts (SPLs), could be explored in future research to provide greater data manipulation flexibility.

*E. Validity of the Findings*

The study retained strong internal and external validity despite its sample size limitation. Blockchain technology helped solve issues of *data integrity* and *research reproducibility*. Furthermore, the ROMA framework's robustness was enhanced by quantitative support of its effectivity in programming roles optimization, confirming previous research findings [3,4,7-10].

The findings' validity was further reinforced by the mixed-methods research approach, combining participant qualitative feedback with experimental quantitative data for a more profound understanding of participants' lived experiences in various human/AI collaborative programming roles.

## VI. Conclusion

This study demonstrates the effective incorporation of blockchain technology for the storing of experimental data, hence improving reproducibility and transparency in empirical software engineering. The dataset, currently being collected through Solana Tx Memos, is accessible to the public at https://bit.ly/bdcga. The results indicate that optimizing programming roles based on personality can markedly enhance individual motivation, performance, and psychological well-being, while also improving team collaboration and productivity.

Notably, the findings suggest that although participation in the Navigator role poses challenges in remote environments, professionals with high *Openness to Experience* are most motivated in the Pilot role.

The tamper-proof Solana blockchain ledger enables ongoing, voluntary inclusion of fresh empirical data, upholding ethical standards of reproducibility and transparency and allowing findings to remain relevant as socio-technical contexts evolve with emerging technologies.

Future research could advance role optimization by exploring complementary personality pairings, allowing individuals to engage more frequently and for extended periods in their preferred roles.

The ROMA framework and its associated tools adapt these human/AI collaboration insights to the needs of VSEs and Gen-Z undergraduates, fostering a flexible, motivating, and adaptive environment. When coupled with smart contracts, this approach holds the potential for adaptive optimization of programming roles across diverse, cross-cultural environments, underscoring its broader applicability for human/AI collaboration.